\documentclass[runningheads]{llncs}
\usepackage[T1]{fontenc}
\usepackage[dvipsnames]{xcolor}
\usepackage{tikz,lipsum,lmodern}
\usepackage[most]{tcolorbox}
\usepackage{comment}
\usepackage{graphicx}
\usepackage{hyperref}
\usepackage{xcolor}

\usepackage{url}

\begin{document}
\title{Anonymity-washing}
%
%

\author{Szilvia Lestyán\inst{1,2}\orcidID{0009-0007-1898-1512},
William Letrone\inst{3}, 
Ludovica Robustelli\inst{3} \\
Gergely Biczók\inst{4}\orcidID{0000-0002-3891-3855
}}
\authorrunning{Sz. Lestyán et al.}
%
\institute{Inria \and
Institut national d'études démographiques, France\\
\email{szilvia.lestyan@inria.fr}\\
\and
DCS, Nantes University, CNRS\\
\email{\{william.letrone,ludovica.robustelli\}@univ-nantes.fr}
\and
CrySyS Lab, Budapest Univ. of Technology and Economics \\
\email{biczok@crysys.hu}}
\maketitle              
\begin{abstract}
Anonymization is a foundational principle of data privacy regulation, yet its practical application remains riddled with ambiguity and inconsistency. This paper introduces the concept of anonymity-washing---the misrepresentation of the anonymity level of ``sanitized'' personal data---as a critical privacy concern. While both legal and technical critiques of anonymization exist, they tend to address isolated aspects of the problem. In contrast, this paper offers a comprehensive overview of the conditions that enable anonymity-washing. It synthesizes fragmented legal interpretations, technical misunderstandings, and outdated regulatory guidance and complements them with a systematic review of national and international resources, including legal cases, data protection authority guidelines, and technical documentation. Our findings reveal a lack of coherent support for practitioners, contributing to the persistent misuse of pseudonymization and obsolete anonymization techniques. We conclude by recommending targeted education, clearer technical guidance, and closer cooperation between regulators, researchers, and industry to bridge the gap between legal norms and technical reality.
\end{abstract}

\section{Introduction}
Anonymization is widely regarded as a crucial tool for protecting privacy in an era of big data processing. Theoretically, it serves as a means to mitigate risks associated with the misuse of personal data by ensuring that individuals can no longer be identified. In practice, however, anonymization remains an imprecise science, often misunderstood and misapplied. Many datasets that are presented as anonymized continue to pose significant re-identification risks due to improper techniques or evolving technological capabilities. This gap between the intended function of anonymization and its real-world implementation has led to growing concerns about \textit{anonymity-washing} — a phenomenon in which organizations claim to have achieved strong privacy protections through anonymization while failing to provide meaningful safeguards. Note that anonymity-washing is a specialized form of \textit{privacy-washing}\footnote{a particularly timely research theme, see \url{https://www.dagstuhl.de/en/seminars/seminar-calendar/seminar-details/25112}}~\cite{cirucci2024oversharing}.

The General Data Protection Regulation (GDPR) establishes anonymization as a mechanism through which personal data can be rendered outside the scope of data protection laws. Recital 26 of the GDPR defines anonymization as the process by which data is \textit{``rendered anonymous in such a manner that the data subject is not or no longer identifiable.''} However, the absence of clear, practical guidance on how to achieve this standard has resulted in inconsistent implementations and legal uncertainties. Many organizations either overestimate the effectiveness of their anonymization processes or struggle to comply due to conflicting regulatory interpretations. Additionally, courts have recognized that anonymization is never absolute — what is considered anonymous\footnote{In this paper we use the word \textit{anonymous} whenever we cite or reference a text that has used the same expression.} today may become identifiable tomorrow as technology advances.  
Despite the importance of anonymization, the regulatory and educational landscape remains fragmented and inadequate. On one end of the spectrum, legal guidelines provide high-level definitions and compliance requirements but lack technical specificity. On the other end, academic research offers rigorous, mathematically grounded approaches to anonymization that are often inaccessible to practitioners who do not have advanced expertise in statistics or computer science. This disconnect has left engineers, data scientists, and policymakers without the necessary tools to implement anonymization effectively. The result is widespread reliance on outdated or insufficient methods---such as k-anonymity and l-diversity---that have been repeatedly shown to fail against modern re-identification attacks~\cite{gadotti2024anonymization}.  

Furthermore, anonymity-washing is exacerbated by inconsistent regulatory interpretations across jurisdictions. The European Union has exercised significant global influence on data privacy regulation, with many countries modelling their laws after the GDPR. However, even within the EU, national data protection authorities and courts have issued conflicting opinions on what constitutes effective anonymization, leading to uncertainty among organizations attempting to comply. Beyond Europe, frameworks such as the United States' de-identification standards under the Health Insurance Portability and Accountability Act (HIPAA) and the California Consumer Privacy Act (CCPA), Japan’s Act on the Protection of Personal Information (APPI), and emerging guidelines such as the Brazilian General Data Protection Law, the \textit{Lei Geral de Proteçao de Dados} (LGPD), further demonstrate that approaches to anonymization lack uniformity at the international level, making cross-border data governance exceedingly complex.  

Another critical factor enabling anonymity-washing is the lack of accessible educational resources for practitioners. Engineers and software developers responsible for implementing anonymization frequently lack adequate training and rely on either high-level legal guidelines or complex, research-oriented papers that do not offer practical guidance. Several regulatory bodies and experts have called for clearer standards, including the European Data Protection Board (EDPB), national data protection authorities (such as the National Commission for Information Technology and Civil Liberties in France (CNIL) and the Federal Commissioner for Data Protection and Freedom of Information (BfDI) in Germany), and research institutions. Yet, despite these calls for action, practitioners continue to report difficulties in accessing concrete, actionable information on how to apply anonymization techniques effectively. 

In light of these challenges, this paper argues that ambiguities in regulatory guidance, outdated technical approaches, and gaps in practitioner education may lead to anonymity-washing.
While prior works have addressed specific aspects of the problem---such as legal critiques of anonymization under data protection law~\cite{rubinstein2016anonymization,stalla2016anonymous,ohm2009broken,burt21anonstandardsguide}, technical limitations of anonymization techniques~\cite{aggarwal2005k,cohen2020towards,langarizadeh2018effectiveness,gadotti2024anonymization,el2015critical,cormode2010minimizing}, or even highlighting key misunderstandings \cite{edpbaepdmisunderstandings}---these contributions offer only a partial view of the broader landscape. In contrast, our work provides a comprehensive analysis of the multiple, interrelated issues underlying anonymity-washing. We expand on the existing literature by integrating a wide range of sources, including legal cases, regulatory interpretations, and technical guidelines, while offering a systematic critique of technical documentation. Furthermore, we provide an international perspective that, to our knowledge, has not been previously compiled in a single work.

First, in Section~\ref{Contextual} we introduce the concept of anonymity-washing and situate it within the broader landscape of privacy discourse. In Section~\ref{overwievguidance}, we examine the legal foundations and the regulatory ambiguity surrounding anonymization. Next, Section~\ref{guidance} presents an overview and critique of technical guidelines and educational resources, highlighting the gaps practitioners face.
Section~\ref{practices} explores the practical implications of anonymity-washing, including legal cases and implementation failures.
Finally, Section~\ref{discussion} offers recommendations and concluding reflections on how to address the risks of anonymity-washing through clearer guidance and improved institutional coordination.


\section{Contextual Elements}
\label{Contextual}
\subsection{Anonymization terminology}

The anonymization landscape is complex, with multiple laws advocating for different requirements. But many points of contention stem from the terminology surrounding the topic of anonymization. To begin, it is interesting to look at the terminology developed by the International Organization for Standardization (ISO), as it constitutes the main standard-setting body with international influence. Several ISO standards touch on the topic of data anonymization. These global standards have recognized the importance of anonymization in various contexts. ISO/IEC 29100:2024(en) establishing a common privacy terminology, defines anonymization as a \\

\begin{center}
\textit{``[A] process by which personally identifiable information (\ldots) is irreversibly altered in such a way that a \textbf{[data subject]} (\ldots) can no longer be identified directly or indirectly, either by the PII controller alone or in collaboration with any other party.''}
\end{center}

The same document defines pseudonymization as a
\begin{center}
\textit{``[A] process applied to personally identifiable information (PII) (3.7) which replaces identifying information with an alias.''}
\end{center}
The other term, ``de-identification'', is usually considered more neutral and broader than anonymization, although sometimes conflated with the latter\cite{chevrier2019use,garfinkel2015identification}. Indeed, according to ISO, ``de-identification'' refers to 
\begin{center}
\textit{``[A] process of removing the association between a set of identifying attributes (3.14) and the data subject (3.4).''}\footnote{ISO/IEC 20889:2018(en) Privacy enhancing data de-identification terminology and classification of techniques. See also, the more recent ISO/IEC 5207:2024(en) Information technology — Data usage — Terminology and use cases.}.
\end{center}It results that anonymization implies the highest degree of privacy, while the more specific process of pseudonymization is a step below anonymization in terms of re-identifiability. In contrast ``de-identification'' is the general term describing the process through which data is made confidential\footnote{Note that the technical terminology could be used slightly differently; here we are discussing only the legal definitions.}.
While some jurisdictions mostly follow the ISO terminology, others, unfortunately, do not~\cite{aliki,Wallace}. An example is the fact that the term ``de-identification'' is not even used within the EU's GDPR, while several important US instruments, such as the HIPAA~\cite{hipaa} and the CCPA~\cite{CCPA} use it in place of anonymization. In the same vein, Nigeria and Malawi's Data Protection Acts do not use the term ``anonymization'', despite referring to both ``de-identification'' and ``pseudonymization'' in their statutes~\cite{aliki}. In contrast, Japan's \textit{Act on the Protection of Personal Information}, much like the EU, does not refer to de-identification. Finally, a cursory look at the relevant literature in social science reveals that authors themselves appear to have subscribed to different terminologies~\cite{gadotti2024anonymization,chevrier2019use}. \\

\noindent Beyond word choice, there seems to be no equivalence between the terms when they are used to refer to data records that have undergone the appropriate treatment to exempt data controllers and processors from their obligations under data protection laws. That is to say, the tolerance level towards identifiability tends to vary across jurisdictions~\cite{aliki}. Discrepancies sometimes exist within a single legal system, as in the US, where re-identifiability tolerance may vary depending on the nature of the data contained in a record, and the projected use of the record~\cite{garfinkel2015identification}. In the EU, the situation is no less confusing, as ``anonymization'' suffers from conflicting interpretations~\cite{el2015critical} (see details in Section~\ref{overwievguidance}).

\begin{tcolorbox}[enhanced,attach boxed title to top center={yshift=-3mm,yshifttext=-1mm},
  colback=black!5!white,colframe=black!50!,colbacktitle=black!30!,
  title=Take-away,fonttitle=\bfseries,
  boxed title style={size=small,colframe=black!50!} ]
  These variations and inconsistencies make it difficult for practitioners to understand and determine the required level of protection, hindering the understanding and adequate application of anonymization techniques. 
\end{tcolorbox}

\subsection{``Anonymity-washing'' as the misrepresentation of actual confidentiality levels}

Due to interpretative instability, the terms that compose the anonymization terminology should not necessarily be taken at face value. Not only are practitioners affected by the confusion in the terminology, but individuals are affected as well, as they may put more trust in information processes than they should. 
In order to better understand this effect, we must look at interpretations of \textit{privacy-washing}. In the course of an analysis on questionable data practices of tech industry giants, Girucci gives the following definition~\cite{cirucci2024oversharing}:
\begin{center}
     \textit{``The purposeful conflation of security with privacy, the disregarding of more granular definitions of privacy (social vs. institutional privacy as well as data types including explicit, implicit, aggregated, and inferred), and a general reliance on offline privacy expectations that are no longer applicable to online spaces.''}. 
\end{center}

\noindent Despite its provocative tone, the term privacy-washing is more than a mere rhetorical device. Indeed, privacy-washing can accurately describe situations where data privacy guarantees deviate from the standards to which the concerned entities purportedly committed. Evidently, the concept of privacy-washing is broad: it can cover a variety of subjects like cybersecurity and third-party data sharing. This paper is focused on privacy-washing in the anonymization context because deceptive privacy representations in this context are highly likely. In fact, while the anonymization vocabulary taken at face value is unambiguous, it does little to convey the actual fragility~\cite{ohm2009broken} of current anonymization methods: 
\begin{center}

\textit{``The way companies and the media talk about de-identified data matters, and data holders regularly play fast and loose with the concept of anonymity. The terms ``anonymous'' and ``anonymization'' simply over-promise. They create expectations of near-perfection and lull people into a false sense of security''}~\cite{rubinstein2016anonymization}.
\end{center}
\noindent Data controllers could be tempted to exploit the complexity within current data privacy terminology to mislead data subjects regarding the safety and confidentiality of their data, resulting in anonymity-washing. In essence, anonymity-washing refers to situations involving the misrepresentation of anonymity levels of a data record. Recently, the Federal Trade Commission (FTC), as the main agency dealing with consumer protection in the US, dealt with anonymity-washing cases. In a recent communication, the FTC warned that unwarranted claims of anonymity could constitute deceptive consumer practices, reiterating that pseudonymous identifiers in the form of hashing do not constitute anonymization, as some businesses have claimed: 
\begin{center}
\textit{``Companies should not act or claim as if hashing personal information renders it anonymized. FTC staff will remain vigilant to ensure companies are following the law and take action when the privacy claims they make are deceptive''}\cite{FTC12}.
\end{center}
Remark, how this highlights the manipulative aspect of anonymity-washing.\footnote{It bears noting that the FTC’s view on hashing is consistent with the practices in the EU, where hashing constitutes a method of pseudonymization that does not suffice on its own, in making data records fall outside the GDPR’s scope, due to the likelihood of privacy harms that may result from sharing the records.} In the EU, potential anonymity-washing cases have been scrutinized by data protection authorities, and some practices have been challenged in Court. For instance, the Italian Data Protection Authority (Garante) recently sanctioned the Italian National Institute of Statistics for its failure to deploy the necessary measures to avoid re-identification of the data it used for statistical analysis. In its order, the Italian authority explained~\cite{GPDP24INI}; 
\begin{center}
    \textit{``Simply having organizational measures or ethical codes is not enough to satisfy data protection principles.''}
\end{center}
In this case, data controllers claimed to have upheld data protection principles while the data subjects remained, in fact, easily re-identifiable from their data records.\\

\noindent The question of intentionality behind deceitful anonymity statements deserves a brief focus, as the term ``washing'' implies an intentional action, motivated by a paucity of resources, time constraints, or uncertain goals regarding projected uses of data records. Except that intention in this context can be difficult to prove. Sometimes, anonymity-washing cases are so blatant that the willingness to deceive leaves no doubt. Other times, anonymity-washing is harder to prove and therefore appears incidental, giving the impression that data controllers and/or processors are acting in good faith while deploying weaker solutions. There is, of course, a risk of mischaracterization. Still, it may never be possible to prove with a high degree of confidence that a data controller and/or processor acted in good faith, since defendants are likely to claim to be acting in good faith when notified, and in the course of legal proceedings. 

\begin{tcolorbox}[enhanced,attach boxed title to top center={yshift=-3mm,yshifttext=-1mm},
  colback=black!5!white,colframe=black!50!,colbacktitle=black!30!,
  title=Take-away,fonttitle=\bfseries,
  boxed title style={size=small,colframe=black!50!} ]
  Anonymity-washing is a subset of privacy-washing, which refers to the misrepresentation of the anonymity level of data. The phenomenon is exacerbated by several factors, including unclear terminology.
\end{tcolorbox}


\section{Overview of regulatory guidance on data anonymization}
\label{overwievguidance}

On 25 July 2024, the European Commission published its second report on the implementation of the GDPR~\cite{gdpr}.
One of the key issues highlighted in the report is the persistence of differing interpretations among national data protection authorities, which undermines the uniform application of the GDPR. This discrepancy gives rise to legal uncertainty; thus, businesses are confronted with divergent administrative requirements across different Member States. In this regard, the Commission seeks to reiterate its request, previously made in 2020~\cite{EC20comm}, to support practitioners by providing clearer guidance and materials to facilitate GDPR compliance. This issue is particularly pertinent in the context of anonymity washing, as the Commission has reported in~\cite{EC24}: 
\begin{center}
    \textit{``Some stakeholders also consider that certain data protection authorities and the Board adopt interpretations that deviate from the risk-based approach of the GDPR, [and] (\ldots) mention as areas of concern: (i) the interpretation of anonymization; (\ldots)''.} [As a result, the report] \textit{``underline}[s] \textit{the need for additional guidelines, in particular on anonymization and pseudonymization (\ldots)''.} 
\end{center}

\subsection{EU regulations}
The abrogation of Directive 95/46/EC~\cite{EC95dir} (Data Protection Directive or DPD) and the adoption of the GDPR did not affect anonymization. This is confirmed by the endorsement of the Working Party 29's (WP29) Opinion 5/2014 on anonymization by the European Data Protection Board (EDPB), 
which is still in the process of preparing an updated version~\cite{stalla23change}. In its Opinion 5/2014 on anonymization~\cite{wp29}, WP29 recalls the ISO definition of anonymization\footnote{ISO 29100:2024} and that the simple removal of identifiers from personal data does not make the anonymization process irreversible. Account should be taken of all``reasonable means''(including computational power and technological evolution) to re-identify anonymous data. These points are addressed by Recital 26 of the GDPR, stating that : 
\begin{center}
\textit{``Account should be taken of all the means reasonably likely to be used, such as singling out, either by the controller or by another person, to identify the natural person directly or indirectly. To ascertain whether means are reasonably likely to be used to identify the natural person, account should be taken of all objective factors, such as the costs of and the amount of time required for identification, taking into consideration the available technology at the time of the processing and technological developments.''}\footnote{The previous directive 95/46/EC, which the WP29 Opinion 5/2014 referred to, was not as detailed. It contained a general reference to reasonable means that are likely to be used to re-identify the data (without including the notion of 'singling out') and relied on codes of conduct to encourage anonymization practices.}     
\end{center}
\noindent Before deciding on an anonymization method, an anonymization test must be performed to evaluate the risks (singling-out, linkability, and inference). WP29 provides an assessment of the guarantees and shortcomings of each technique from the two main families of anonymization (generalization and randomization) based on these risks. However, several research papers have shown that the analyses provided by WP29 have weak points, and they do not consider these techniques valid~\cite{cohen2020towards,whitehouse14,narayanan2014no,de2016response,stadler2022search,narayanan2019robust}\footnote{For example, the k-anonymity technique does not prevent the risk of singling out, contrary to the conclusions of WP29.}.\\

\noindent More recent guidelines have been adopted by the EDPB on issues related to data anonymization, but they do not contain additional advice. On 17 December 2024, the EDPB published guidelines on the anonymization of AI models~\cite{edpb24AI}. It states that whenever models are trained on personal data, they cannot be considered anonymous. The reason is that many studies on these models have demonstrated their capacity to ``regurgitate'' part of their training datasets~\cite{ayyamperumal2024current}. The EDPB stated that a model is considered anonymous only when, based on appropriate documentation, personal data cannot be inferred either directly (by statistical inference, including the probabilistic functioning of the model) or indirectly (within a user's prompt). If the risk of ``regurgitation'' of personal data persists, a deeper analysis is needed\footnote{The following aspects are required for verification: the source of the data, their preparation and minimization, the training method, the analysis of the model, the resistance of the model to cyber attacks, and the documentation provided.}. 

\noindent Sénéchal criticizes the lack of a threshold of the risk of ``regurgitating'' personal data and the lack of a distinction between the different AI models in these guidelines in~\cite{senechal24avisedpb} (for example, the general-purpose AI models~\cite{ep24reg1689} 
and the ones posing systemic risks. 
This is problematic since anonymization is difficult to implement, especially with unstructured data~\cite{weitzenboeck2022gdpr} 
, which are essentially used to train general-purpose AI models~\cite{wolff2023lessons}. Additionally, the question of whether the data can be separated from the model remains unanswered. It is also not known whether the anonymization of the model implies that of the data it contains. \\

\noindent The EDPB also adopted on 16 January 2025, guidelines on pseudonymization~\cite{edpb25pseudo}, in which it recalls the GDPR definition set out in article 4(5). The Board stressed that, although pseudonymization secures data, whenever the re-attribution of data to a natural person (by linking pseudonyms to additional data) remains possible, the GDPR applies. It recalls that even if the original data are deleted, pseudonymized data become anonymous only if all requirements are met. It is interesting to note that the \textit{guidelines do not provide further information on anonymization requirements}. This is regrettable for two reasons: first, updated guidelines on anonymization have yet to be issued; and second, as the Spanish Data Protection Authority has pointed out, confusion between anonymization and pseudonymization remains a common misunderstanding among data controllers~\cite{edpbaepdmisunderstandings}. 
\begin{tcolorbox}[enhanced,attach boxed title to top center={yshift=-3mm,yshifttext=-1mm},
  colback=black!5!white,colframe=black!50!,colbacktitle=black!30!,
  title=Take-away,fonttitle=\bfseries,
  boxed title style={size=small,colframe=black!50!} ]
  Guidelines on anonymization need to be updated (as they have not been since 2014). The information provided by the EDPB on pseudonymization and anonymization of AI models does not resolve the contradictions of its previous guidelines and the practical difficulties controllers are confronted with when implementing anonymization protocols in real life.
\end{tcolorbox}

\subsection{Anonymization regimes beyond the EU}
The uncertainties resulting from the changing interpretation within the EU undermine the so-called ``Brussels effect'', when non-EU states take inspiration from the EU's laws for building their own legal regime. Data flows often involve entities located in different jurisdictions, including non-EU countries~\cite{HoJoo2023Comparison,sardor}. 
Moreover, data protection laws usually have some extraterritorial effects, which means that multiple regimes are sometimes applicable simultaneously. Hence, it is vital to ensure that legal regimes on anonymization do not contradict each other. Yet, a survey conducted by the OECD in 2019 found that 
\begin{center}
\textit{``uncertainty regarding legal privacy regimes''} and \textit{``incompatibility of legal regimes''} topped the list of the main challenges to cross-border data flows~\cite{oecddataflow}. 
\end{center}


\noindent Anonymization guidelines are present in data protection regimes across the globe. There are differences, however, in the approaches and the overall granularity levels exhibited by the relevant frameworks. Notably, some data protection regimes, such as in Japan~\cite{Japanact2003} and the US~\cite{hipaa}, come with relatively detailed guidance on how to achieve the expected levels of anonymization and how to handle the data~\cite{Japan17guidance,Japan20guide}. Concerning data transfer between the EU and the USA, the previous Privacy Shield, which was adopted on the basis of the European Commission's decision that the USA's level of personal data protection was equivalent to that of the EU, ~\cite{privacyshield16}, was replaced by a revised Privacy Framework, after the EU Court of Justice overturned it~\cite{EUCJc31118Schrems}
In contrast, other regulatory frameworks, such as in Brazil\cite{lgpd}, are not particularly prescriptive and require additional input. 
At the same time, several jurisdictions have initiated efforts to modernize their approaches to data protection, including anonymization. For example, Brazil's national data protection authority, the ANDP, is set to clarify what measures could be implemented to ensure anonymity in accordance with its LGPD in the upcoming years~\cite{ANPDagenda2526}. A call for public participation in that effect has been published in early 2024. In 2023, the Data Security Council of India (DSCI) published a roadmap considering possible orientations for a national data anonymization regime~\cite{DSCI}. 
At the intra-state level, in Québec, the \textit{Regulation respecting the anonymization of personal information} was published in 2024~\cite{Quebecreg}. The text is very prescriptive and seeks to clarify the distinction between anonymity and pseudonymity, aligning with the EU's view. In the EU, the EDPB is expected to publish new guidelines on anonymization later this year. It is expected that the new guidelines will fix the inconsistencies introduced by the WP29 Opinion 5/2014 on anonymization, thereby clarifying the dominant approach at the EU level~\cite{el2015critical}. \\

\noindent Anonymization is still a maturing field. Valuable guidelines on anonymization are often released after the publication of the main body of law. Hence, recent data protection laws such as China's \textit{Personal Information Protection Law}, or India's \textit{Digital Personal Data Protection Act}, will need to be complemented with guidelines on anonymization~\cite{stalla2025identifiability,HoJoo2023Comparison}. Furthermore, while precise anonymization parameters are still not consistent across jurisdictions, the basic premises of anonymization law remain the same; anonymization levels may vary, and so are the obligations placed upon data handlers~\cite{HoJoo2023Comparison}. \\

\noindent Whatever the approach, it seems that regulators are left with two choices: either leaving enough leeway for data handlers to determine for themselves which methods and policies would meet their expectations, or prescribing exactly which technical and organizational measures would meet their expectations. Both approaches have their merits and shortcomings. On the one hand, there is an inherent limitation on the degree of granularity that can be achieved in the law. Excessively precise regulations and guidelines may pose problems at the implementation stage and may prove to be overly restrictive, as has already been seen with the WP 29 Opinion 5/2014. The limited technical knowledge of regulators may constrain the formulation of highly detailed guidelines anyway. On the other hand, too much leeway could seriously undermine the purpose of data protection laws by increasing the likelihood that poorly anonymized data records will fall outside their scope.  

\begin{tcolorbox}[enhanced,attach boxed title to top center={yshift=-3mm,yshifttext=-1mm},
  colback=black!5!white,colframe=black!50!,colbacktitle=black!30!,
  title=Take-away,fonttitle=\bfseries,
  boxed title style={size=small,colframe=black!50!} ]
Anonymization laws seem to have been evolving independently, with differing requirements and definitions. Whether all the ambiguities will be fixed and whether every actor will converge around the same interpretation remains to be seen.
\end{tcolorbox}

\section{Overview and Critique of Guidelines}
\label{guidance}
\subsection{Contradictory guidelines and uncertain standards within the EU}

Some authors consider that the EU Data Protection Law lacks a clear definition of anonymization~\cite{burt21anonstandardsguide}. Unlike pseudonymization, the GDPR fails to define anonymization in its Article 4 titled``definitions''. The reason is that anonymization and pseudonymization are considered in a binary way\footnote{\href{https://techcrunch.com/2017/10/07/how-anonymous-wifi-data-can-still-be-a-privacy-risk/}{https://techcrunch.com/2017/10/07/how-anonymous-wifi-data-can-still-be-a-privacy-risk/}}. Indeed,  anonymized data is excluded from the scope of the GDPR, while pseudonymized data is entirely subject to it. This is unlike some legislation (such as in Japan), which suggests a lighter regime for pseudonymized data while still imposing obligations on anonymized data. The EU approach fails to acknowledge that there is always a risk of re-identification with anonymisation, and that pseudonymized data may require lighter protection than 'classic' personal data. This is despite WP 29 acknowledging this risk. On this point, another Opinion of WP29 on the concept of personal data issued in 2007~\cite{wp2907personaldata} clarified the difference between anonymization and pseudonymization. Recalling ISO's previous definitions, they explained that anonymization protects privacy, while pseudonymization represents a technical, reversible process. Nevertheless, data can still be considered anonymous, even when re-identification remains possible, but complementary measures to prevent re-identification are implemented. This flexible approach was not supported in the Opinion 5/2014 on anonymization~\cite{wp29}, which applies together with the previous Opinion on the concept of personal data. In Opinion 5/2014, WP29 required the aggregation of data (into group statistics) and the destruction of raw data (identifiers) to ensure correct anonymization. Nevertheless, the objective still remained to prove that the likelihood of re-identification was negligible. On this point, the Commission's guidelines~\cite{EC19comm} on the free flow of non-personal data suggest that it is often difficult to assess the effectiveness of an anonymization procedure. Indeed, besides many academic papers~\cite{altmanacm,gadotti2024anonymization,de2016response,narayanan2019robust,aggarwal2005k,de2013unique}, even a study commissioned by the European Parliament’s ITRE Committee has shown that it is possible to re-identify supposedly anonymized data~\cite{EP17itre}.\\ 
The risk of re-identifying anonymous data stems from the technical limitations of anonymization and the lack of clear and realistic guidelines on the subject. This suggests that re-identification is not only a consequence of poor anonymization implementation. 

\noindent Moreover, national DPAs disagree on how to implement anonymization. The French DPA, CNIL, adopts the WP29’s approach to anonymization; other national DPAs are more flexible. For example, the UK's ICO (Information Commissioner’s Office), despite UK's own data protection regulation being largely similar to that of the EU, states that~\cite{ICOrisk}:
\begin{center}
\textit{``The DPA does not require anonymization to be completely risk free---you must be able to mitigate the risk of identification until it is remote. If the risk of identification is reasonably likely, the information should be regarded as personal data--- (\ldots). Clearly, 100\% anonymization is the most desirable position, and in some cases, this is possible, but it is not the test the DPA requires.''}, 
\end{center}
and Ireland's DPC (Data Protection Commission) writes\cite{DPCpseudovsanon19}:
\begin{center}
\textit{``Organisations don’t have to be able to prove that it is impossible for any data subject to be identified in order for an anonymization technique to be considered successful. Rather, if it can be shown that it is unlikely that a data subject will be identified given the circumstances of the individual case and the state of technology, the data can be considered anonymous.''}
\end{center}
The CNIL’s guidelines on anonymization are of particular interest, as they establish a strict standard to determine whether data can be considered anonymous. They assert that data is anonymous only when it is \textit{impossible} to re-identify the data subject. However, they recognise that when the risks of singling out, linkability, and inference are not met, data can be deemed anonymous if a subsequent analysis indicates a negligible risk of re-identification~\cite{CNILanon20}. Unfortunately, the definition of what constitutes a \textit{negligible} risk remains ambiguous, since anonymization is a context-dependent process, depending on the nature of the data and its intended use.  \\
\begin{tcolorbox}[enhanced,attach boxed title to top center={yshift=-3mm,yshifttext=-1mm},
  colback=black!5!white,colframe=black!50!,colbacktitle=black!30!,
  title=Take-away,fonttitle=\bfseries,
  boxed title style={size=small,colframe=black!50!} ]
The lack of a clear, harmonized definition of anonymization across EU legal texts and among national authorities aggravates the complexities of anonymization and favors the confusion between pseudonymization and anonymization. Although anonymization and pseudonymization rely on similar techniques, they have different aims: anonymization aims to conceal the data subject's identity, whereas pseudonymization aims to protect privacy by making re-identification more complex but not impossible. Technical documents should help practitioners assess when to opt for anonymization rather than pseudonymization. However, most of the available technical guidance does not address this issue.
\end{tcolorbox}

\subsection{Technical Documents}

While non-technical guidance often lacks the precision needed for implementation, technical documentation is not always more helpful. In several cases, companies have claimed they could not find clear guidance on how to anonymize data—a claim sometimes countered by DPAs pointing to existing documents~\cite{CNIL24cegedim}. However, our review shows that most guidelines are often hard to find (not available or ignored by DPAs) or not practically useful (entry-level).
We reviewed the websites of the five most active EU DPAs (France, Austria, Ireland, Germany, and Italy). Most do not provide detailed technical materials:
\begin{itemize}
    \item CNIL (France) offers introductory guides~\cite{CNILanon19,CNILanon20}, repeating WP29 content, and a clear (though potentially misleading) explanation of pseudonymization~\cite{CNILresearchpseudoanon22}.
    \item Garante (Italy) provides an overview~\cite{GPDPguide23} to implement the GDPR, and mainly reiterates previous legal guidelines.
    \item BfDI (Germany) has policy papers and speeches~\cite{BfDItalk22,BfDIaktuelle23,BfDItalk24,BfDItalk25}, but limited technical depth.~\cite{burkert2020positionspapier} focuses on the importance of anonymization,~\cite{BfDIarbeitsptelem23} and~\cite{BfDIarbeitspsmart22} discuss risks of other issues related to personal data.
    \item DSB (Austria) offers legal advice only.
    \item DPC (Ireland) stands out with well-structured and clear guidance~\cite{DPCbasics19,DPCpseudovsanon19} on legal questions; however, it does not offer practical advice. 
\end{itemize}

\noindent Outside the EU, the UK’s ICO provides excellent guidance, including on state-of-the-art methods like differential privacy and other PETs~\cite{ICOanonguide22,ICOpets23}. However, some examples are oversimplified or technically wrong\footnote{See the case study on differentially private mixed noise addition}

\noindent The UK Anonymization Network (UKAN) also offers practical tools, such as a decision-making framework that uniquely addresses attacker modelling~\cite{UKAN20_guide}. However, most chapters remain general (entry-level), in contrast to the referenced DIS method that requires Master's degree-level statistical knowledge.

\noindent The anonymization guide of Singapore's PDPC~\cite{singaporeguide22} is accessible and educative, guides the reader from data discovery to risk measures, giving informative examples; however, it is also entry-level and adds little beyond other existing material.

\noindent Some statistical agencies provide additional resources. The National Institute of Statistics and Economic Studies in France (INSEE) offers slides and working papers~\cite{insee19risquereid,insee24anon}, but most lack practical detail. A notable exception is Bergeat’s work~\cite{inseesdc2016} comparing and explaining experiments done by two anonymization software tools: $\mu$-Argus and SDCMicro. It also gives plenty of citations, however, only to sources on statistics (no computer science references). It is aimed at statisticians, and it could serve as a continuation to other introductory materials, but the reader who already has at least a Bachelor's degree in mathematics or statistics.
Other guides, such as~\cite{inseeguide24}, focus on confidentiality rules within the French statistical service rather than techniques.
Statistical documents only mention statistical tools and use a different language from that of computer scientists. This could, unfortunately, result in ignoring some state-of-the-art methods, such as differential privacy. A good example is the working paper~\cite{jmsutopie} that details the anonymization process applied to a large French administrative database where the authors experimented with different methods including k-anonymity, all-m anonymity, and l-diversity. They mention that they have tried to apply DP; however, they abandoned the experiment due to a lack of expertise.

\noindent Academic papers are another option, but they often assume advanced statistical or mathematical knowledge (Master's or PhD level), making them inaccessible to many practitioners. Moreover, choosing appropriate methods from the literature is difficult without deep expertise, which may explain the frequent use of outdated or misapplied techniques in practice~\cite{insee24anon,inseesdc2016,jmsagri,jmsutopie}. Some expert-written materials aimed at non-technical readers exist~\cite{wood2018differential,nguyen2020techniques,nguyen2014techniques}; however, they are rarely cited in public or institutional guidance.

\subsubsection{Books}

Books on anonymization tend to target either high-level management (e.g.~\cite{craig2011privacy,sharma2019data,nissenbaum2009privacy}) or technical researchers. Some, like~\cite{bhajaria2022data}, cover a broad range of privacy topics but lack methods for evaluating anonymization quality. Jarmul's work~\cite{jarmul2023practical} offers a more hands-on perspective, including differential privacy and privacy engineering workflows, making it useful for practitioners. Stallings' book~\cite {stallings2019information} is a strong general-purpose resource, well-suited for short training programs. 
\begin{tcolorbox}[enhanced,attach boxed title to top center={yshift=-3mm,yshifttext=-1mm},
  colback=black!5!white,colframe=black!50!,colbacktitle=black!30!,
  title=Take-away,fonttitle=\bfseries,
  boxed title style={size=small,colframe=black!50!} ]

Most technical anonymization resources are either too simplistic or too involved, offering little practical use for professionals. Practical regulatory guidance is rare and often legalistic. This leaves practitioners with a fragmented landscape, outdated methods, and an incentive to abandon anonymization altogether. Bridging these gaps requires targeted, accessible, and technically sound educational materials.
\end{tcolorbox}

\section{Inadequate Practices}

\label{practices}
EU case law lacks clarity with regard to anonymization practices. The most relevant cases focus on clarifying the concept of personal data. However, the interpretations provided help to assess what an anonymized dataset is \textit{not}. Moreover, the Court's application of Recital 26 to real cases provides valuable insight into the question of whether data remain anonymous despite a residual risk of re-identification. The following subsections focus on relevant case law concerning the notion of personal data, the implications of which are important for anonymization, and inadequate anonymization practices, demonstrating a lack of awareness on the subject both at the institutional and organizational levels. The final subsection also addresses the inadequate practices that arise from the confusion between anonymization and pseudonymization. At least one example is given for each subsection; they refer to the most representative and recent cases, but these are not exhaustive of all the existing case law on the subject.     

\subsection{Case-law on personal data}

The General Court's SRB vs EDPS decision is a good example of a decision relating to personal data whose implications are also relevant for anonymization~\cite{EUGCsrbedps}. SRB (Single Resolution Board) carried out an insolvency procedure against Banco Popular. Within this procedure, some data were processed to assess the eligibility of the participants for compensation. Each participant was identifiable by means of an alphanumeric code generated randomly. The staff processing these data only had access to these codes and not to the key identifying them. The EDPS (European Data Protection Supervisor) considered these data pseudonymized~\cite{EC18reg}, but its decision was challenged before the General Court. Using a risk-based approach, the General Court decided that the data were anonymous. Indeed, according to  Breyer's Court of Justice case law\cite{EUCJ16Breyer}, the additional information (the key) needed to re-identify the data subjects remained inaccessible to the processing staff. The fact that the staff could not legally access the complementary data that would allow re-identification proved enough to consider that no \textit{reasonable means} existed to re-identify the data, which thus remained non-personal~\cite{EUGCsrbedps}. 

This example shows that the EU Court of justice's (EUCJ) case law on personal data builds upon its precedents rather than undergoing a radical evolution. On this point, Breyer's decision~\cite{EUCJ16Breyer} was about the dynamic nature of IP addresses. These IP addresses are subject to change with each connection. The plaintiff initiated legal proceedings against the Republic of Germany for its practices concerning the storage and registration of these data. The Court of Justice had to determine whether dynamic IP addresses should be considered personal data for the service provider. The Court decided that the retention of all information by a single individual was not a prerequisite for data being considered as personal. This meant that a third party could retain such re-identifying information, and that this circumstance did not affect the qualification of the data. However, the Court acknowledged that an assessment was necessary to determine the reasonableness of combining this information, taking into account the effort, time, and cost associated with the operation, as well as the accessibility of this additional information (enabling user identification) to the service provider. Given the legal restrictions on such access in Germany, the Court determined that in the absence of legal means to obtain this information, the data in question were not deemed personal. The doctrine posits that two fundamental elements have been applied since the Breyer decision to ascertain the personal nature of data. These are: (1) the distinguishability of the data, defined as the capacity of the data points to identify an individual, and (2) the availability of additional data to ``situationally relevant entities'' who are capable of associating these data with a physical person\cite{stalla2025identifiability}.
 
The Scania decision perfectly~\cite{EUCJ23Gesamtverband} illustrates this methodology. In this case, the Court was asked to determine the legal status of a vehicle identification number (VIN), a unique alphanumeric code assigned by manufacturers to identify the proprietor of a vehicle. In its decision, the Court stated that the VIN can be personal data for independent operators and vehicle manufacturers if the former have the additional data that enable re-identification, and for the latter if they make the VIN available. 
The availability of data is considered in conjunction with the capacity of isolating owners of vehicles or all other people who have a title on them. Some authors have suggested an evolution in the interpretation of personal data, attributing this change to the Court's categorization of independent operators and manufacturers as ``situationally relevant entities'', capable of associating VIN with additional identifying information. 

\begin{tcolorbox}[enhanced,attach boxed title to top center={yshift=-3mm,yshifttext=-1mm},
  colback=black!5!white,colframe=black!50!,colbacktitle=black!30!,
  title=Take-away,fonttitle=\bfseries,
  boxed title style={size=small,colframe=black!50!} ]
The consistency of the Court's jurisprudence on the concept of personal data is paramount to contrast the phenomenon of anonymity-washing. This is particularly crucial given the occurrence of poor anonymization or privacy-protecting practices, which the EUCJ is entitled to sanction.
\end{tcolorbox}

\subsection{Case-law on inadequate practices}

A relevant case law that sanctioned the European Commission~\cite{EUCJ24gocec} demonstrates the importance of taking into account publicly available data to assess the risk of re-identification. The applicant received European funding as a researcher. The funds had been misappropriated, and the costs were ordered to be reimbursed. The Commission published a press release summarizing the decision, without mentioning the applicant's direct identifiers to protect their privacy. However, the researcher brought an action for the annulment of the press release, since it contained identifiable data. The General Court dismissed it, and the applicant appealed its decision to the EUCJ. The Court considered that \textit{``information relating to the gender of a person who is the subject of a press release, that person’s nationality, his or her father’s occupation, the amount of the grant for a scientific project and the geographical location of the entity hosting that scientific project, taken together, contain information that may allow the person who is the subject of that press release to be identified, in particular by those working in the same scientific field and familiar with that person’s professional background''} and goes on that this circumstance, \textit{`` does not allow the risk of identification of the data subject to be regarded as insignificant.''}. Although the judgment focuses more on privacy protection than anonymization, it highlights the lack of understanding that simply deleting direct identifiers is insufficient to minimize the risk of re-identification. This point is also relevant to anonymization. 

Confusion about anonymization practices is also widespread among companies. A good example is the IAB Europe case-law~\cite{EUCJ24IAB}. The company established a set of guidelines aimed at ensuring compliance with the GDPR concerning the collection of browsing data via a TC String (a series of characters coding the user's preferences). This string, which the company claimed to contain anonymized data, could later be used by companies for commercial purposes. The Court ruled that the TC String was a form of personal data, given its capacity to allow individuals to be identified by associating it with additional information (such as an IP address). Despite third parties retaining this additional information, IAB Europe was able to obtain it. 

In this regard, the relationship between the EUCJ case-law, employing a risk-based approach to assess the reasonable means likely to be used to re-identify the data, and the WP 29 Opinion 5/2014 on anonymization appears complex~\cite{stalla2016anonymous}. On the one hand, discordance persists in the discourse of WP 29 between the zero-risk approach (re-identification must be negligible if not impossible) and the necessity for reasonableness, given that all anonymization techniques are considered imperfect~\cite {el2015critical}. On the other hand, the need to destroy raw data and to aggregate them in order to achieve anonymization, as required by WP 29, is not met by the case law of the EUCJ, which considers data to be anonymous even if the original data are not deleted, the only relevant aspect being the impossibility (legal rather than technical) to access the additional data that enable re-identification. These contradictions contribute to privacy washing practices by making it difficult to distinguish the company's bad faith from its lack of knowledge about anonymization methods, especially when data are processed for commercial purposes. In the case of IAB Europe, for example, the company assumed that, due to the unavailability of additional information, TC String contained anonymized data. This mistake is frequently observed among firms that find it difficult to distinguish anonymization from pseudonymization and anything in between. This assertion is supported by some decisions adopted by the national DPAs.

\subsection{Confusions arising from the difference between pseudo- and anonymization}

One of the most relevant decisions on anonymization dates to 5 September 2024, namely the CEGEDIM SANTE case~\cite{CNIL24cegedim}. It perfectly illustrates the interactions between the WP 29 guidelines on anonymization, the EUCJ's case-law, and the context-dependent nature of anonymization. Indeed, the nature of the data (health data) and their use (creation of a health data repository) are not neutral. CEGEDIM designs and sells secretarial software for the medical sector. The company collected patient health data from doctors who agreed to participate in creating a health data repository. These data were allegedly anonymized with k-anonymity techniques. The decision was based on two key factors: (1) the WP29 test \footnote{Linkability, re-identification, and inference.}, which determines whether the data were anonymized, and (2) EUCJ case law. However, the rapporteur isolated a 6-year-old patient with a medical condition, which would suggest pseudonymization, unless re-identification is proved to be ``negligible'' by ``reasonable means''. On this point, the CNIL concluded that the available data could be easily re-identified. However, the company contended that there were few educational materials on anonymization and that the guidelines lacked precision, rendering them unsuitable for legal certainty. Despite CNIL's rejection of the complaint, the WP 29 Guidelines on anonymization are not updated concerning the current risks associated with k-anonymity. Furthermore, the CNIL has not specified what constitutes a 'negligible risk'. This observation suggests that anonymity-washing may not be a deliberate practice.

\begin{tcolorbox}[enhanced,attach boxed title to top center={yshift=-3mm,yshifttext=-1mm},
  colback=black!5!white,colframe=black!50!,colbacktitle=black!30!,
  title=Take-away,fonttitle=\bfseries,
  boxed title style={size=small,colframe=black!50!} ]
Despite the consistency of the EUCJ case-law relating to personal data, confusion is widespread at the institutional level and among firms as far as the distinction between personal and non-personal data. Such confusion originates from the difficulty in implementing anonymization, which is a probabilistic and context-dependent process, and in distinguishing it from pseudonymization and anything halfway between these two techniques. This issue could enable unintentional anonymity-washing.
\end{tcolorbox}
\section{Discussion and overture}
\label{discussion}
The gap between regulatory guidance and technical solutions gives space to anonymity-washing, whether intentional or unintentional. While frameworks exist to support anonymization efforts, their inconsistent application, misinterpretation, and continued reliance on outdated methods — repeatedly shown to be ineffective — often create a false sense of compliance and security. Several key issues contribute to this phenomenon.

In Section~\ref{overwievguidance} we explain the \textbf{lack of clear guidance} to apply regulations and definitions.
Guidelines lack a clear definition of pseudonymization and omit a definition of anonymization. This inconsistency can be a source of confusion and anonymity-washing. Nevertheless, establishing a coherent terminology is not a straightforward task, as personal data can be strongly situation-dependent, and data can be used in numerous ways.

We have also shown that \textbf{guidelines are} often \textbf{outdated or unreliable}. It has been shown that the anonymization techniques in the WP 29 Opinion 5/2014 on anonymization are no longer reliable, and relevant questions persist, given the contradictory nature of the document.
Consequently, guidelines on anonymization remain to be updated, and the information provided by the EDPB on pseudonymization and anonymization of the AI models does not address this gap.

Next, we have seen that the \textbf{differing interpretations} of the regulations among authorities compromise their uniform application, leading to legal uncertainty for businesses and organisations. For example, some authorities, like the CNIL, adopt a stringent approach, while others, like the ICO and the DPC, are more flexible.

Furthermore, in Section~\ref{practices} we show that there is a \textbf{lack of awareness} among practitioners. They often do not recognize that their data can constitute personal data, leading to a failure to implement privacy by design principles~\cite{BfDIarbeitsptelem23}. Such misconceptions could be eliminated by adequate training and guidance that would give the right tools to practitioners and engineers to be able to competently assess their datasets and apply anonymization methods.

We believe that one crucial cause of this shortcoming is the \textbf{lack of understanding} of anonymization methods. Many practitioners, particularly those without advanced mathematical skills, struggle to understand and apply fundamental privacy principles. The complexity of privacy-enhancing technologies (PETs) creates an additional barrier, making it difficult for non-specialists to implement effective anonymization (e.g.:~\cite{BfDIarbeitspsmart22,jmsutopie,EUCJ24IAB}).

Another consequence of this educational deficiency is that organizations continue to \textbf{rely on outdated anonymization methods}, such as k-anonymity and l-diversity, despite their well-documented vulnerabilities~\cite{gadotti2024anonymization}. This reliance stems from a lack of awareness regarding modern privacy-preserving techniques, as well as limited resources for evaluating and adopting alternative approaches.

However, this \textit{lag} between the state-of-the-art and the most popular, but outdated tools is neither newfound nor unparalleled. There has always been a \textbf{collaboration gap} between academia and industry that limits the transfer of theoretical advancements into practice. Without structured mechanisms to facilitate knowledge-sharing, industry professionals may not only struggle to integrate the latest research into their anonymization strategies but also completely ignore it.

\subsection{Overture}

To address these issues and mitigate the risks of anonymity-washing, we would like to raise awareness among privacy experts and encourage them to facilitate adoption by practitioners for practitioners. With this objective in mind, we suggest the following actions:

In order to clarify the complexity of anonymization as a probabilistic and context-dependent process we believe that one of the most important action to take is to \textbf{develop a comprehensive anonymization curriculum} that could be promoted and distributed by data protection authorities either in the form of training programs offered or thorough and up-to-date educational resources, such as books and hands-on exercises aided with structured guidance on fundamental privacy principles and their real-world applications. Key components should include: 
(1) A clear explanation of privacy threats and their manifestations in various datasets. (2) An overview of widely accepted privacy definitions (besides k-anonymity and l-diversity, adding differential privacy and cryptographic methods), including their advantages and drawbacks. (3) How these techniques can defend against said privacy threats. (4) Techniques for evaluating privacy technologies and applying them to real-world scenarios. (5) Strategies for auditing privacy risks and implementing mitigation measures in large-scale datasets. (6) Best practices for integrating privacy considerations into broader software engineering projects. (7) Case studies illustrating data re-identification and the consequences of inadequate privacy protections.
(8) Adding hands-on learning resources, such as Jupyter notebooks and real-world datasets.

Moreover, educational resources should be \textbf{tailored to diverse audiences}. Given the varying levels of expertise among practitioners, privacy education must be designed to accommodate both technical and non-technical professionals. It is also important to emphasize the \textbf{use of state-of-the-art methods}. Practitioners should be trained to critically assess privacy techniques and understand the limitations of traditional methods in modern, large datasets. To be sufficiently critical, we should also include adequate privacy risk assessments based on attacker capabilities, data sensitivity, and intended data usage. Finally, we believe that a curriculum of this depth can not be delivered without \textbf{enhancing the collaboration between academia and industry}. We acknowledge that finding a common language between academia and industry is not always straightforward. It takes time and effort: joint initiatives, workshops, and training programs should be encouraged to bridge the gap between theoretical advancements and practical implementation. By realizing these recommendations, organizations and policymakers can move beyond superficial compliance efforts and work toward fostering a robust, meaningful approach to data privacy that will help reduce the prevalence of anonymity-washing and ensure that anonymization practices align with contemporary privacy risks, industrial demands, and capacities.\\

We would like to emphasize that we are not suggesting procedural safeguards or checklists, nor do we mean to imply that clear guidance alone can resolve the complexities of anonymisation. Instead, we propose addressing the underlying technical limitations and misunderstandings. Technical guidelines should not be directed at policymakers, but rather at engineers, technical personnel, or even social scientists who need to anonymise data themselves. 

\subsection{Future work}
As a support of the collaboration between academia and industry, we wish to create a repository of guidelines and educational materials using the accumulated knowledge that we have used to write this paper. We envision constructing a proper website 
aided with instructions that would help practitioners navigate and find the appropriate guideline, document, or educational resource to their needs.

Furthermore, we conjecture that there is one more potential source of many of the aforementioned problems, namely, the use of popular anonymization tools. Thus, we have already started examining these existing tools; firstly, to corroborate our conjecture, and secondly, to be able to properly include them in the aforementioned website, equipping practitioners with the necessary understanding and comparison of these products.

\subsubsection*{Acknowledgments. } This work has been partly funded by Project no. 138903, implemented with the support provided by the Ministry of Innovation and Technology from the National Research, Development, and Innovation Fund, financed under the FK\_21 funding scheme.

\subsubsection*{Disclosure of Interests.} The authors declare no competing interests. Project funding is acknowledged in the Acknowledgments section.

\bibliographystyle{splncs04}
\bibliography{bibliography}

\end{document}